\documentclass[11pt,twoside]{article}

\usepackage{asp2006}
\usepackage{fancyhdr,graphicx,caption,rotating,multirow,lscape}
\DeclareGraphicsExtensions{.eps,.eps.gz,.ps,.jpg,.tiff}
\begin{document}

%%% Fill in title
\title{A Search for Transiting Hot Planets as Small as Neptune in the Open Cluster M37}  

%\title{}

%%% Fill in author names, use initials and surname and affiliation
%%% One author
%\author{J. W. Doe}
%\affil{Max-Planck Institut f\"ur Astronomie, 69117 Heidelberg, Germany}

%%% Two authors and more
\author{J. D. Hartman}
\affil{Harvard Smithsonian Center for Astrophysics, 60 Garden, Cambridge, MA~02138, USA}
\author{B. S. Gaudi}  
\affil{Department of Astronomy, The Ohio State University, Columbus, OH~43210, USA}
\author{M. J. Holman}
\affil{Harvard Smithsonian Center for Astrophysics, 60 Garden, Cambridge, MA~02138, USA}
\author{B. A. McLeod}
\affil{Harvard Smithsonian Center for Astrophysics, 60 Garden, Cambridge, MA~02138, USA}
\author{K. Z. Stanek}  
\affil{Department of Astronomy, The Ohio State University, Columbus, OH~43210, USA}
\author{J. Barranco}
\affil{Harvard Smithsonian Center for Astrophysics, 60 Garden, Cambridge, MA~02138, USA}
%\affil{International Institute for Extra-solar Planets, Moon Station}    
%\author{J. W. Doe3}
%\affil{International Institute for Extrasolar Planets, Moon Station} 

%\author{}
%\affil{}

\begin{abstract} %%% Abstract to run on from here.
We are conducting a transit survey of the open cluster M37 using the
Megacam instrument on the 6.5 m Multiple-Mirror Telescope. We have
obtained $\sim$ 4500 images of this cluster over 18.5 nights and have
achieved the precision necessary to detect planets smaller than
Saturn. In this presentation we provide an overview of the
project, describe the ongoing data reduction/analysis and present some
of our preliminary results.
\end{abstract}

%%% MAIN BODY OF TEXT GOES HERE. 
%%% Invited Talks : 10 pages
%%% Talks         : 6  pages
%%% Posters       : 3  pages
%%%IF YOU FEEL THAT THE INFORMATION PROVIDED 
%%% IN THIS FILE IS NOT SUFFICIENT, CONSULT THE FILE aspauthor2006.pdf. IT
%%% CONTAINS "INSTRUCTIONS FOR AUTHORS USING LATEX2E MARKUP", 
%%% SECTIONS 2.3-2.6 FOR HELP WITH EQUATIONS, FIGURES, AND TABLES.

\section{Introduction}   %%% Top level section head (remove "%" symbol)

One can separate surveys for transiting planets along two axes: surveys of bright stars vs. surveys of faint stars and galactic field surveys vs. surveys of stellar clusters. Each type of survey has its own advantages and disadvantages so that there are a wide range of possible transit survey strategies each with different scientific goals in mind \citep{Charbonneau.06}.  This presentation describes a deep survey of an open cluster: M37 (NGC 2099).

The advantage of surveying a stellar cluster is that the stellar population in the survey can be well characterized. As a result, it is straightforward to determine the planet frequency (or an upper limit) as functions of stellar and planetary mass and orbital period from the number of transiting planets that are discovered. Knowing this frequency would provide a fundamental constraint for theories of the formation and evolution of planetary systems. Moreover, knowing the properties of the stars in the survey has the advantage of lessening the probability of spending valuable time following-up false positive transit detections, a problem which has plagued field surveys. For these reasons there have been a number of surveys for transiting planets in stellar clusters. This includes surveys of globular clusters \citep{Gilliland.00, Weldrake.05} and open clusters. The latter includes UStAPS \citep{Street.03, Bramich.05, Hood.05}, EXPLORE-OC \citep{vonBraun.05}, PISCES \citep{Mochejska.05, Mochejska.06}, STEPSS \citep{Burke.06} and MONITOR \citep{Aigrain.06}.  

Despite a large number of surveys, to date no confirmed transiting planet has been discovered in a stellar cluster. So far the open cluster surveys have only provided upper limits on the planet frequency, all of which lie above the frequency inferred from the other surveys.  The fundamental problem facing an open cluster survey is that very few rich open clusters exist. Moreover, the richest clusters, which have at most a few thousand members, all lie more than a kilo-parsec away.  Surveys of open clusters are necessarily deep surveys for which the expected number of planet discoveries is low (typically one or two planets).

\citet{Pepper.05} developed a formalism for estimating the expected yield of a cluster survey.  This work has given the insight that the minimum radius planet that can be detected around a cluster star is approximately constant for all stars that have a photometric precision limited by source shot noise. For stars with light curves that do not have correlated noise the minimum detectable radius is inversely proportional to the square root of the telescope diameter. \citet{Pepper.06} have noted that it may be possible to find planets as small as Neptune, or even Super Earths using existing ground based telescopes and instruments. While radial velocity and micro-lensing surveys have recently uncovered several planets in this regime \citep{Butler.04, Santos.04, McArthur.04, Beaulieu.06, Lovis.06} there is a substantial motivation to discover one that transits its host star. Moreover, the frequency of planets in this regime is essentially unknown, thus even an upper limit on this frequency would represent a significant improvement in our knowledge of these planets. Finally, the discovery of even a single transiting planet in a stellar cluster would provide a unique opportunity to determine the mass and radius of the planet with a precision that is not limited by uncertainties in the mass and radius of the star.

Following this motivation we have undertaken an ambitious twenty night survey of the open cluster M37 (NGC 2099) using the Megacam wide-field mosaic CCD camera on the 6.5m Multiple-Mirror Telescope (MMT). By using a significantly larger telescope than has heretofore been utilized for a transit survey we are able to achieve a precision that should in principle allow us to detect planets as small as Neptune. This survey also provides a unique opportunity to study low-amplitude variability in an intermediate age open cluster. In \S 2 we describe the planning, observations and data reduction. We briefly describe a few preliminary results in \S 3.

\section{The Survey}
\subsection{Planning}
To demonstrate the feasibility of a full scale transit survey with the MMT we conducted a preliminary three night survey of the open cluster NGC 6791 \citep{Hartman.05}. We were able to achieve a point-to-point precision of better than 1 mmag for the brightest stars; the noise was not dominated by systematics. These results suggested that for a closer open cluster we could achieve the precision necessary to detect planets as small as Neptune which would provide a 1 mmag deep transit around a solar radius star (with a deeper transit for smaller stars). 

To select the open cluster for a transit survey we applied the formalism of \citet{Pepper.05} to all open clusters visible from the MMT in December/January. We chose this timing constraint so that observations could be obtained continuously over two trimesters during the long winter nights. Data for the open clusters was obtained from WEBDA where available with a literature survey to determine the cluster mass function for the most promising candidates. Besides maximizing the formal expected number of detections, we also considered the age, metallicity, angular size and presence of additional data for the clusters. Among the top choices we chose M37 (05:52:19, +32:33:12) for its solar metallicity ([Fe/H] = +0.09, \citet{Mermilliod.96}) and intermediate age (520 Myr, \citet{Kalirai.01}). 

\subsection{Observations}
The observations were obtained over twenty four nights (eight of which
were half nights) between December 21, 2005 and January 21, 2006. We
obtained a total of $\sim$ 4500 images. We used the Megacam instrument
\citep{McLeod.00} which is a $24\arcmin$ x $24\arcmin$ mosaic
consisting of 36 2k$\times$4k, thinned, backside-illuminated CCDs that
are each read out by two amplifiers. We used an $r^{\prime}$ filter to
maximize the sensitivity to smaller stars while avoiding fringing that
would occur at longer wavelengths. The mosaic has a pixel scale of
$0.08\arcsec$ which allows for a well sampled point-spread-function
(PSF) even under the best seeing conditions. This camera provides a
field of view large enough to contain the cluster without
dithering. And, because of its fine sampling, one can obtain $2\times
10^7$ photons in $1\arcsec$ seeing from a single star prior to
saturation setting the photon limit for the precision in a single
exposure to 0.25 mmag.

The average time between exposures (including read-out and time spent initializing for the next exposure) was 24 seconds. Ideally one would like to observe the same stars from exposure to exposure while spending as little time as necessary with the shutter closed. In other words, one would like to maximize the exposure times. There are, however, competing factors that limit how long one can expose. As the exposure time is increased more stars are lost to saturation, and, more importantly, the fraction of the image that is lost to saturated stars and artifacts (diffraction speaks and bleed columns) increases. Not only does this decrease the number of stars that can be observed, it also compromises the quality of the image and the ability to reduce the image to achieve high precision photometry for any of the stars. To determine the optimal exposure time we obtained a series of preliminary observations of M37 during a Megacam engineering run on October 29, 2005. We found that in $1\arcsec$ seeing an exposure time of sixty seconds was the longest we could go before saturation posed a problem for image quality. To keep the same stars saturated in each image we then varied the exposure time as the seeing and atmospheric extinction changed.

\subsection{Data Reduction}
When read out with 2$\times$2 binning, a single mosaic image requires 189 megabytes of storage. The entire run produced nearly a terabyte of data which presents a challenge for data reduction. We analyze each chip independently, so there are more than 160,000 images to reduce. To save i/o time we performed the preliminary CCD corrections, including bias correction and flat-fielding, using our own software. We constructed a single master flat-field from twilight sky flats taken over the course of the run. This was done, in part, because conditions were acceptable for flat-fielding on only a handful of evenings.

To perform photometry we used the image subtraction technique due to \citet{Alard.98} and \citet{Alard.00}, using a slightly modified version of Alard's {\scshape Isis 2.1} package. We constructed a reference image for each chip from $\sim 100$ of the best seeing images (the actual number used varied from chip to chip). Stars are identified on the reference and each science image using {\scshape Sextractor} 2.3.2 \citep{Bertin.96}, we used {\scshape Sextractor} rather than the \emph{extract} routine in {\scshape Isis} as {\scshape Sextractor} appeared to handle blended stars better and also performed better on poorer seeing images (we still run \emph{extract} to remove cosmic-rays). The reference star list is then matched to the image star list using a third order polynomial, we then transform the higher signal to noise reference image to the object image. For images with better than 1$\arcsec$ seeing we used a Gaussian convolution kernel to broaden them before performing subtraction. We then perform subtraction adjusting the FWHM of the Gaussians used in the kernel according to the value expected from the difference in seeing between the reference and object images.  To determine the differential photometry on the difference images we performed aperture photometry, scaling the result by the aperture sum of the PSF convolved with the kernel. We performed PSF fitting photometry on the reference images using {\scshape Daophot} version 2 \citep{Stetson.87, Stetson.92} to convert the differential light curves into magnitudes. To ensure proper flux scaling we used the PSF from {\scshape Daophot} as input to the differential photometry routine.

The above procedure produced light curves for 24,000 sources with  $14 < r < 25$ . The light curves are then sent through the following pipeline to identify transits:
\begin{enumerate}
\item Remove images from the light curves that are outliers in more than a given fraction of all the light curves. The cutoff is chosen for each chip based on a visual inspection of the histogram of the fraction of light curves that have a given image as a three-sigma outlier.
\item Remove 0.9972696 day period signals from the light curves. This is done to remove artifacts due to, for example, rotating diffraction spikes that have a period of exactly 1 sidereal day. The signal is removed by binning the light curves in phase and then adjusting the points in a bin by an offset so that the average of the bin is equal to the average of the light curve.
\item Find periodic signals in the light curves using the Lomb-Scargle algorithm \citep{Lomb.76, Scargle.82, Press.89}, remove significant detections using a low-order Fourier series. This is done to remove continuous semi-periodic variations due to, for example, star spots.
\item Detrend the light curves using the TFA algorithm \citep{Kovacs.05}. The trend list for each light curve consists of the other stars on the chip with root-mean-square (RMS) $< 0.1$ that are well outside the photometric radius of the star in question. There are typically $\sim 250$ stars in the trend list for each chip.
\item Search for transits using the BLS algorithm \citep{Kovacs.02}.
\end{enumerate}

\section{Preliminary Results}

\begin{figure}[h]
\centering
\begin{minipage}[c]{.49\textwidth}
\includegraphics[width=5cm]{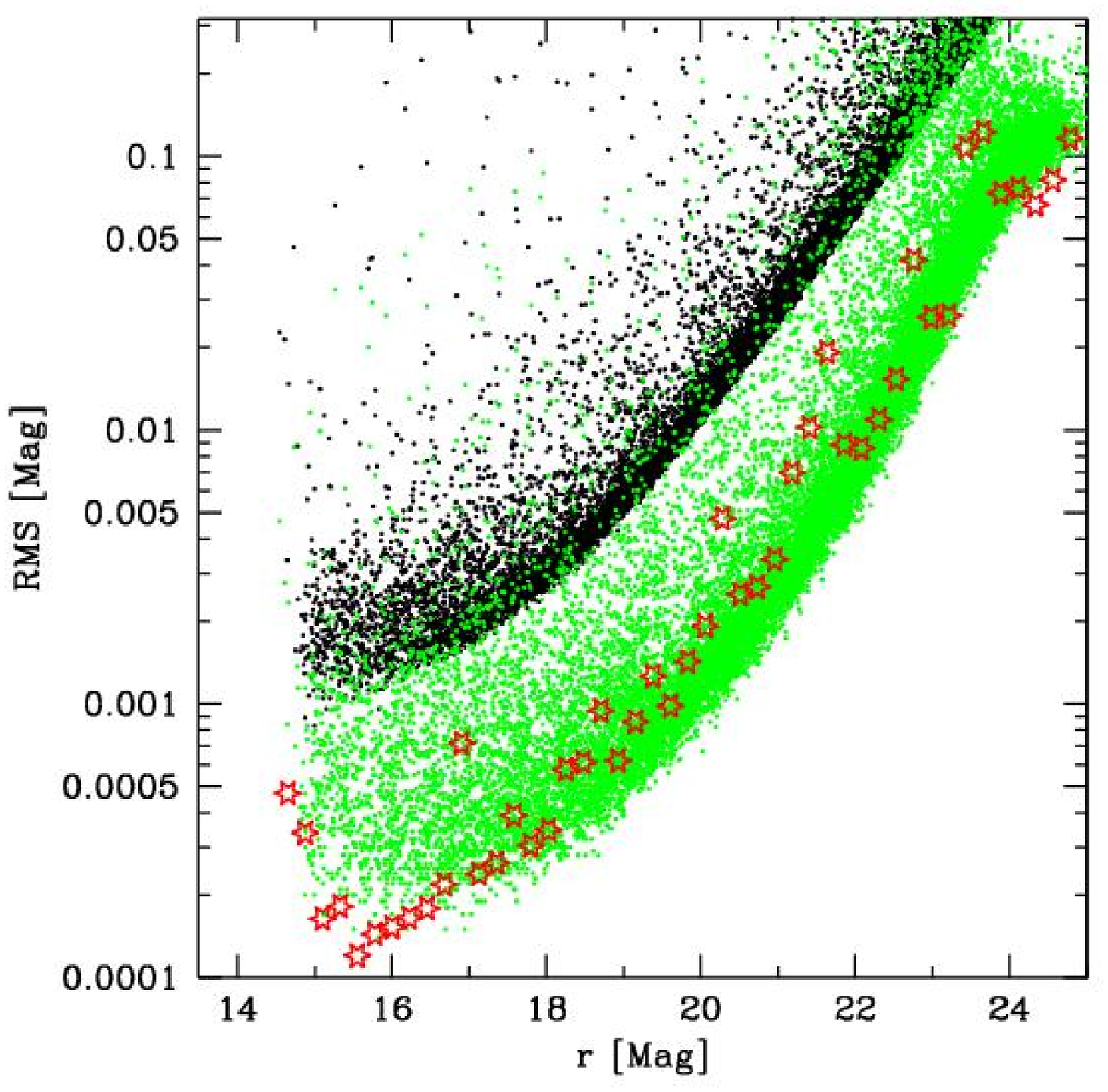}
\end{minipage}
\begin{minipage}[c]{.49\textwidth}
\includegraphics[width=5cm]{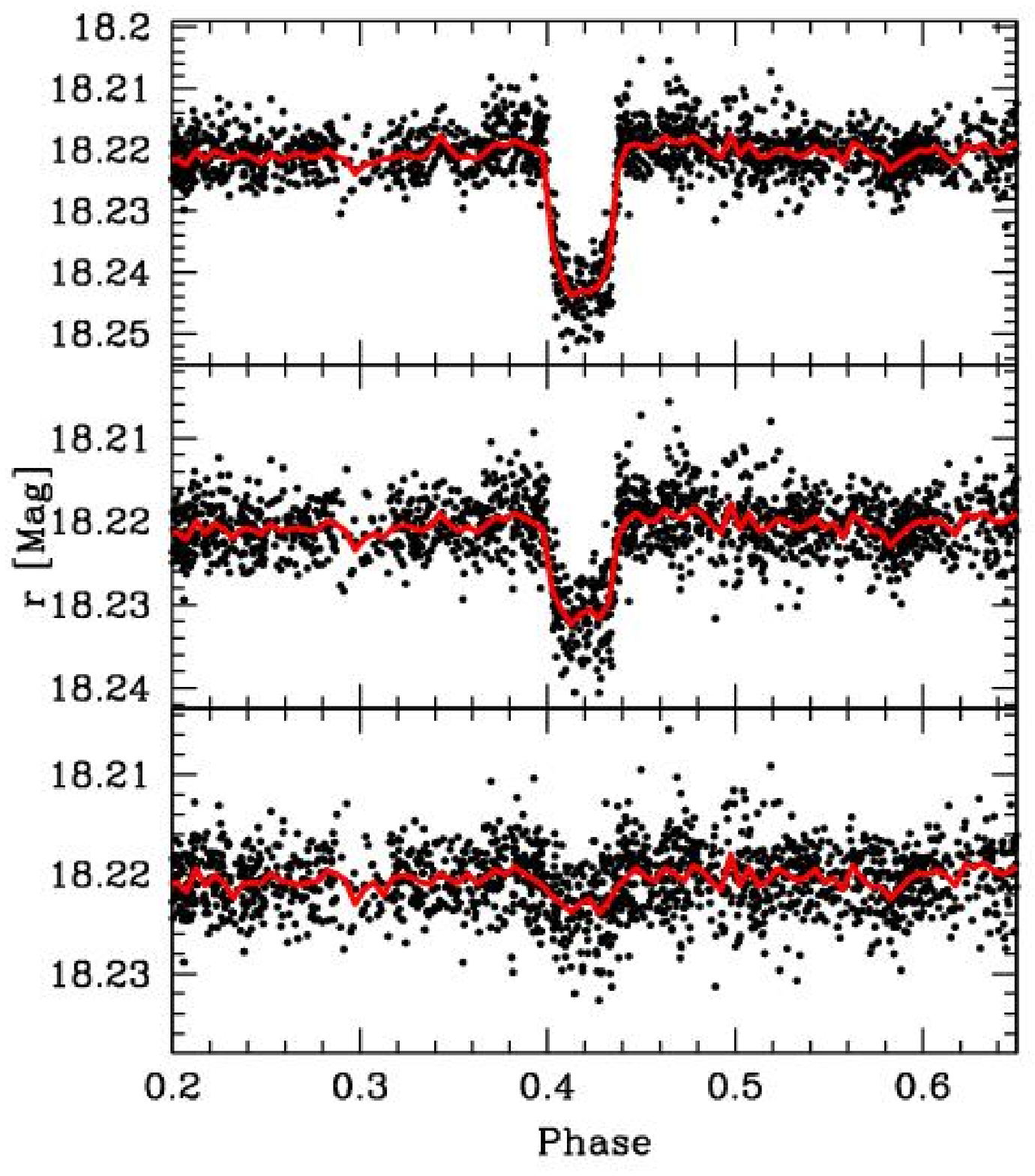}
\end{minipage}
\caption{Left: RMS over the entire run as a function of magnitude for stars in M37. Dark points are after detrending with TFA. Light points are after detrending with TFA and applying a moving mean filter with a two hour time-scale. The stars show the expected binned RMS assuming the unfiltered light curves do not have time-correlated noise. Right: light curves of injected transiting planets ranging in size from Neptune (bottom) to Jupiter (top). We are marginally sensitive to planets as small as Neptune, but can easily detect larger planets.\label{fig}}
\end{figure}

As a demonstration of our photometric precision, Fig.~\ref{fig} shows the RMS of the resulting light curves after detrending. We also show the RMS after applying a moving mean filter of two hour width to the light curves - it is apparent that the limit at which time-correlated noise begins to dominate the light curves is well below 1 mmag. As a further demonstration, Fig.~\ref{fig} also shows light curves for injected transiting planets ranging in size from Neptune to Jupiter. While we have marginal detection capability for planets as small as Neptune, transiting planets smaller than Saturn should be easily detected if they exist. Using the above pipeline we have identified a number of candidate transiting planets, spectroscopic follow-up of these systems to rule out false positive scenarios is underway.  Other results, including studies of the variability and rotation of stars in M37 will be presented elsewhere.

%%% FIGURES
%%% Pathnames : Please keep all your figures files in the same 
%%%             directory as the tex file to avoid pathname problems
%%% Colors    : ASP Conference Series books are printed in black 
%%%             and white. Thus please provide a caption compatible
%%%             for black and white prints

\acknowledgements %%% Text of acknowledgments runs on after this command.
This research has made use of the WEBDA database, operated at the Institute for Astronomy of the University of Vienna. 

%%% THE BIBLIOGRAPHY

\end{document}